# Nonlinear optical properties of two and three-dimensional hybrid perovskite using single beam F-scan technique at 1064 nm

J. Serna,[1,a)] J. I. Uribe,[2,3] E. Rueda,[4] D. Ramírez,[2] F. Jaramillo,[2] J. Osorio,[3] and H. García,[5]

[1]*Grupo de Óptica y Espectroscopía, Centro de Ciencia Básica, Universidad Pontifícia Bolivariana, Cq. 1 No 70-01, Campus Laureles, Medellín, Colombia*

[2]*Centro de Investigación, Innovación y Desarrollo de Materiales – CIDEMAT, Facultad de Ingeniería, Universidad de Antioquia U de A, Calle 70 No. 52-21, Medellín, Colombia*

[3]*Grupo de Estado Sólido, Instituto de Física, Universidad de Antioquia U de A, Calle 70 No. 52-21, Medellín, Colombia*

[4]*Grupo de Óptica y Fotónica, Instituto de Física, Universidad de Antioquia U de A, Calle 70 No. 52-21, Medellín, Colombia*

[5]*Department of Physics, Southern Illinois University, Edwardsville, Illinois, USA*

In this paper two-photon absorption coefficient $\beta$ and nonlinear refractive index $n_2$ have been measured at 1064 nm for a new two-dimensional $(CH_3(CH_2)_2NH_3)_2(CH_3NH_3)_2Pb_3Br_{10}$ hybrid perovskite. The nonlinear optical properties were measured using transmission and reflection F-scan techniques. As a reference, a three-dimensional $CH_3NH_3PbBr_3$ has been used. As expected, the values obtained for $n_2$ and $\beta$ show a difference between these two materials. The figure of merit for both, three dimensional and two-dimensional perovskites are $F = 0.31$ and $F = 0.04$, respectively, implying that these materials can be used in photonics applications such as optical limiting devices and all-optical switching.

Lead halide perovskites have emerged in the last years as very attractive materials due to their particular optical and transport properties[1,2], leading to the fabrication of high performance solar cells and more recently to light emitting diodes (LEDs)[3,4] and batteries[5]. The typical structure of a three-dimensional (3D) perovskite is $ABX_3$ being A an organic molecule such as methyl ammonium $CH_3NH_3$ (MA) or formadimidium (FA), or an inorganic cation such as cesium (Cs). Lead (Pb) is commonly used in the B position, while I, Br and Cl are used as X site anions. Combinations of these different atoms have been used to tune the band gap and the stability of the material[6]. Several reports of the non-linear optical properties of this material in thin films have been done[7–9], showing large third-order optical nonlinearity. As a variation of the standard 3D lead halide perovskite, two-dimensional (2D) perovskites have been

---

[a)] Author to whom correspondence should be addressed. Electronic mail: juan.sernar@upb.edu.co

developed[10]. This kind of structures are very interesting light harvesting and emitting materials as the small MA of the 3D perovskite cation is partially replaced by a larger ammonium cations (A') used as a spacer material, thus confining the perovskite in two dimensions because of steric effects[11,12]. This kind of structure with a general formula (A')$_2$(MA)$_{n-1}$PbI$_{3n+1}$ (Ruddlesden–Popper phase) is expected to have different optoelectronic properties. The structure is composed by MAPb layers separated by the A' organic cation that insulates the layers and gives a closed structure, which make 2D perovskites notably moisture-resistant. In this paper, we studied the nonlinear properties of a recently synthesized 2D (CH$_3$(CH$_2$)$_2$NH$_3$)$_2$(CH$_3$NH$_3$)$_2$Pb$_3$Br$_{10}$ perovskite [5,13], simplified as (PA)$_2$(MA)$_2$Pb$_3$Br$_{10}$, where the spacer propyl-ammonium (CH$_3$(CH$_2$)$_2$NH$_3$) is expected to change the absorption band edge and improve the light emission yield [14–17]. We also used CH$_3$NH$_3$PbBr$_3$ (MAPbBr$_3$) as a 3D reference material to compare our findings. Specifically, the nonlinear refractive index $n_2$ and two-photon absorption coefficient (TPA) $\beta$ of both, 2D and 3D perovskites were investigated, using the TF-scan [18,19] and RF-scan technique[20–24]. These techniques, which are based in the well-known Z-scan technique[25], used an electronically focus tunable lens (EFTL) as a dynamic system to generate the nonlinear optical effects in the material. For TF-scan, the light that is transmitted through the sample is collected by a T-integrating sphere and analyzed to obtain the value of $\beta$. In the RF-scan, the light that is analyzed corresponds to light reflected by the surface of the sample, and in this case, the value of $n_2$ is determined. We found through TF-scan and RF-scan configurations, that 2D and 3D perovskite have high nonlinearities, making it a good candidate in photonics applications.

For this work, films were prepared according to the previous reported method[5]. Briefly, Methylammonium iodide (MAI) and methylammonium bromide (MABr), n-propylammonium iodide bromide (PABr) from Dyesol were used as organic cations. Dimethyl sulfoxide (DMSO, Sigma Aldrich) and N,N-dimethylformamide (DMF, Alfa Aesar) were used as solvents, lead iodide (PbI$_2$) (Sigma Aldrich) and lead bromide (PbBr$_2$) (Alfa Aesar) as lead source. In order to obtain the MAPbI$_3$ precursor solution, 159 mg MAI, 461 mg PbI$_2$ and 71.05 µL DMSO (1:1:1 molar ratio) were dissolved in 560 µL N,N-dimethylformamide (DMF, Alfa Aesar). For the rest of the compositions, MABr and PbBr$_2$ were used to obtain stoichiometric MAPbI$_2$Br, MAPbIBr$_2$ and MAPbIBr$_3$ by replacing MAI and PbI$_2$, respectively. For the 2D layered perovskite (CH$_3$NH$_3$)$_2$(CH$_3$(CH$_2$)$_2$NH$_3$)$_2$Pb$_3$Br$_{10}$, PABr, MABr and PbBr2 (2:2:3 molar ratio) were dissolved with 71.05 µL DMSO in 560 µL DMF. Precursor solution was deposited on top of 2.5 cm x 2.5 cm glass substrates by spin coating at 4000 rpm for 25 s. After 10 s of spin coating, 500 µL of diethyl ether was dropped in order to quickly remove the DMF. The films were then annealed at 65 ºC for 1 min plus 100 ºC for 10 min.

To characterize the films, AFM was performed in a MFP-3D AFM (Asylum Research). The tips used were Silicon Ti-Ir coated (Asyelec-01) with nominal spring constant of 2.89 N/m and resonance frequency of 71.7 kHz. Photoluminescence (PL) and Absorption measurements were taken in a Cary eclipse and Cary 100 Varian respectively. X-ray difractograms were collected from obtained powders in a PANalytical difractometer. The samples were scanned from 2θ = 10° to 60° in a Bragg-Brentano geometry, using Cu-K (1.5408 A) radiation with a step size of 0.04 ° and a speed of 5 ° min$^{-1}$.

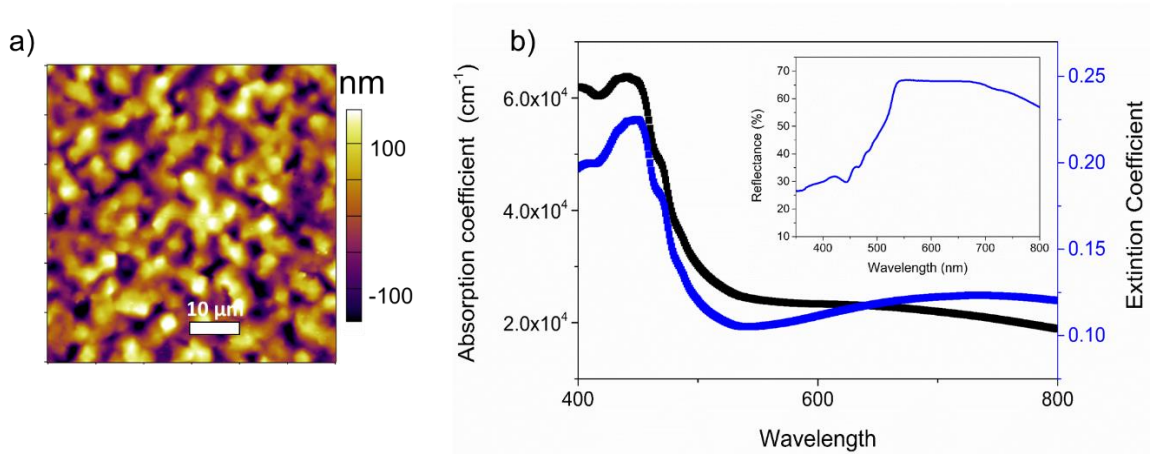

FIG 1. a) AFM image and b) absorption coefficient (black), extinction coefficient (blue) and inset reflectance of the 2D perovskite film. Wavelength units: nm.

Figure 1a shows an AFM image of the 2D film, which presented a granular structure with a grain size around 1 μm and a roughness of 30 nm. This particular morphology allows the adequate measurements of the optical properties of the films without large amounts of scattering. Absorption and extinction coefficients are shown in Fig. 1b. Clearly high values of both are reached for wavelengths below 520 nm, particularly in the absorption values over $10^4$ cm$^{-1}$ are significant for photovoltaic applications. The values near to high wavelengths can be attributed to scattering due to roughness of the film. The inset presents the value for diffuse reflectance. By extrapolating the linear region of the absorption coefficient the electronic energy gap is estimated to be 2.31 eV and 2.36 eV for 3D and 2D perovskite, respectively.

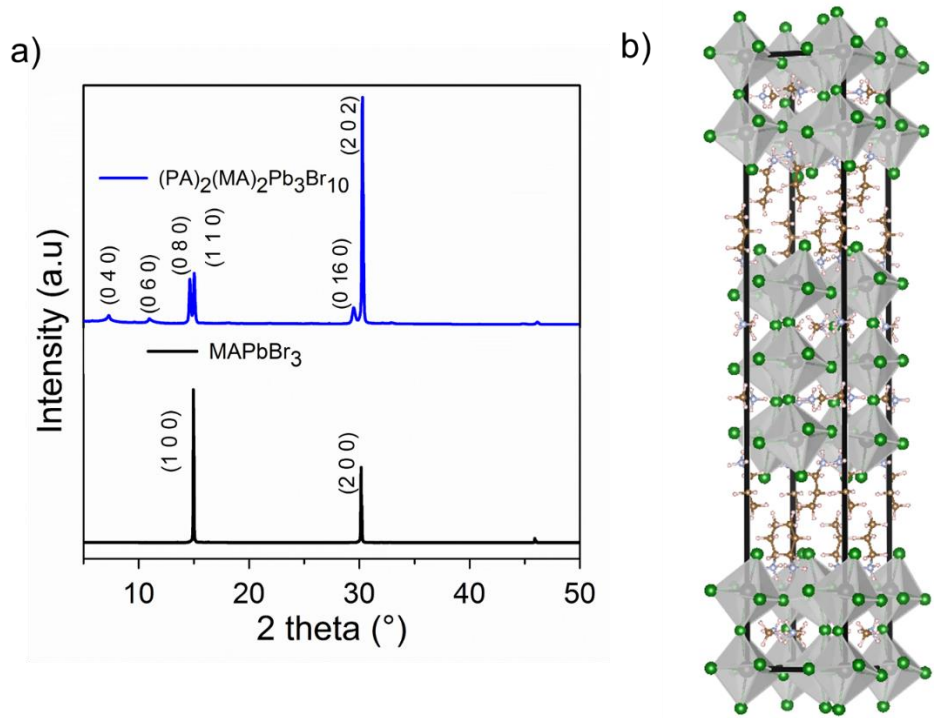

FIG 2. a) XRD diffractogram of the 3D and 2D perovskite films. b) Crystal structure of the 2D perovskite.

To characterize the perovskite-thin films, X-Ray diffraction pattern (XRD) for $(MA)_2(PA)_2Pb_3Br_{10}$ and $MAPbBr_3$ were used, and the results are shown in Fig. 2a. Both XRD patterns match those already reported for these two perovskite films[5]. The obtained 3D perovskite presented a cubic with (100) and (200) diffraction peaks, while the 2D layered structure had more intense diffraction peaks of the (202) plane, indicating preferential growing of the films, which should have consequences in the optical properties. In Fig. 2b is clearly observed the 2D configuration of the material, with planes of 2 octahedrons separated by the PA molecule.

The F-scan experimental set-up is depicted in Fig. 3. In this experiment, the signal is recovered by transmission and by reflection. In the first case, the light that impinges on the sample is transmitted and information about TPA coefficient is obtained. In the second case, the light reflected by the sample surface is analyzed, and information of the nonlinear refractive index is obtained.

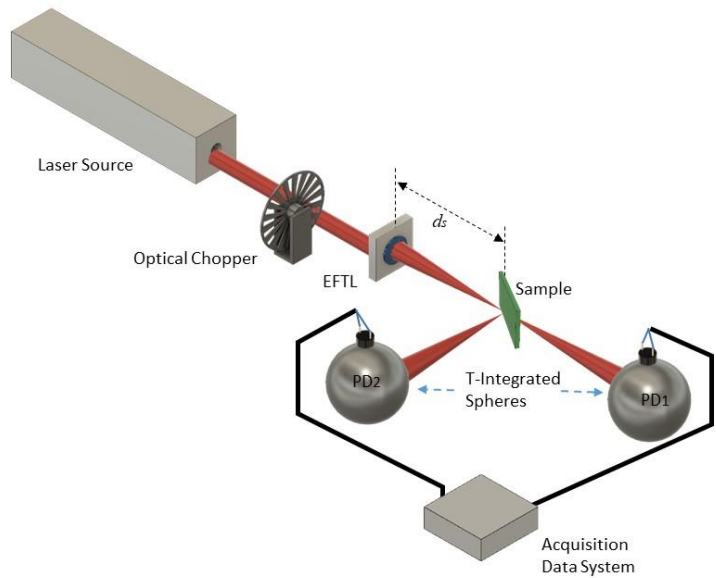

FIG 3. Experimental setup for F-scan configuration. The light that is transmitted by the sample is collected in the T-integrated sphere PD$_1$ (TF-scan). The reflected light is analyzed by PD$_2$ (RF-scan).

A laser Gaussian beam modulated with a chopper impinges on the EFTL, which is a lens that has the capability to vary its focal distance $f$ over a specific range when an electric current is applied to it. The EFTL then focuses the Gaussian beam at different positions. The sample is placed at a fixed position $d_s$ inside the range of the EFTL. The light transmitted through the sample is collected by a T-integrated sphere that has a photodetector PD$_1$, and the light reflected by the sample is collected by a second T-Integrated sphere and a photodetector PD$_2$. These configurations, i.e., transmission and reflection, are named TF-scan and RF-scan respectively. In both cases, the output signal is filtered with a Lock-in amplifier and processed with a computer (PC).

For the experimental implementation of the TF-scan and RF-scan we used a Q-switched laser with repetition rate of 11.4 kHz, pulse width of < 1 ns, and laser emission centered at 1064 nm. The average power at the entrance surface of the sample was 30 mW, and the beam diameter at the EFTL was $D = 1.5 \pm 0.1$ mm. The Electrically Focus-Tunable Lens is an OPTOTUNE-1030, controlled by an OPTOTUNE lens driver that gives a maximum current of 300 mA with a resolution of 0.1 mA, given a focal length resolution of 0.017 mm. We used a large area Si-photodiode to measure the transmitted and reflected laser light. The current generated by the photodiode is sent to a STANFORD RESEARCH 830 dual channel Lock-in amplifier, controlled through a GPIB interface.

To determine the TPA coefficient $\beta$ and nonlinear refractive index $n_2$, we measured the transmittance and reflectance of the nonlinear medium as a function of the focal length, $f$, (see Fig. 3). When the distance $|d_s - f|$ is large the normalized signal has a value close to unity because linear optical effects are produced in the sample. In contrast, small values of $|d_s - f|$ imply that the laser beam is focused near the sample, thus increasing the optical intensity and generating nonlinear optical phenomena such as TPA and changes in the refractive index. Both, TPA and nonlinear refractive index are obtained by fitting theoretical curves given by Eq. (1) and Eq. (4), to the experimental data, using $\beta$ and $n_2$ as the fitting parameters.

For the case of nonlinear optical properties, we shown TF-scan traces for 3D and 2D thin-film perovskites in Fig. 4. In this case, we used glass as substrate such it does not generate nonlinear optical. A normalized transmission peak shows that saturable absorption is presented, therefore a negative sign of TPA is obtained. To calculate the value of $\beta$ we fitted the transmission $T(f)$ experimental data by means of equation (1),

$$T(f) = \frac{1}{B(f)} \int_0^\infty \ln[1 + B(f)\text{sech}^2(\rho)]d\rho, \qquad (1)$$

Where $B(f) = \beta(1-R)I_o(f)L_{eff}$. Here $R$ is the reflection coefficient, $I_o$ is the peak intensity of the beam as the function of the focal length, $L_{eff} = (1 - e^{-\alpha L})/\alpha$ is the effective sample thickness, with $L$ is the sample thickness and $\alpha = 10^{-4} \text{cm}^{-1}$ is the linear absorption coefficient. Finally, $\rho$ is an integration variable expressed as $\rho = 2\ln(1 + \sqrt{2})/\tau$, and $\tau$ is the full width at half-maximum pulse duration. As an important aspect, we found that 3D perovskite presents a $\beta = (-6.0 \pm 2.0)\,\text{cm}/\text{MW}$ which is lower than the TPA coefficient for the 2D configuration, $\beta = (-25.6 \pm 7.3)\,\text{cm}/\text{MW}$.

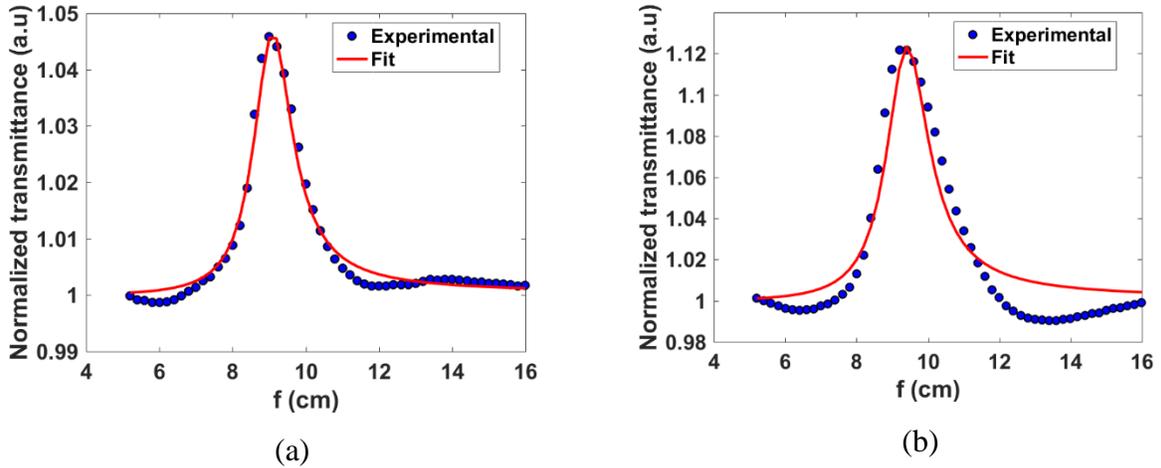

FIG 4. TF-scan results for a) 3D Perovskite and b) 2D Perovskite.

For the case of nonlinear refractive index, we implemented the RF-scan technique, where the sample is tilted an angle $\theta$ with respect to the incident beam. Light reflected by the sample surface is redirected to a T-integrated sphere and the reflectance of the system is analyzed. When the intensity of the beam is high, i.e., when the sample position is closed to the EFTL focal distance, a change in the refractive index is induced into the sample. This situation modifies the intensity of the beam registered by the detector, and the non-linear refractive index $n_2$ can be measured. The intensity of the laser beam reflected by the sample can be calculated using a Fresnel expression for the amplitude reflection coefficient $R(\bar{n},\theta)$. Here $\bar{n} = \bar{n}_0 + \Delta \bar{n}(I)$ is the complex refractive index of the material, with $\bar{n}_o$ being the linear refractive index and $\Delta \bar{n} = n_2 + i\kappa_2$ is the complex nonlinear refractive index[22]. In the last expression $n_2$ is the nonlinear refractive index and $k_2$ is the nonlinear extinction coefficient. The normalized reflectance $\mathcal{R}_N$, measured by a detector coupled to T-integrated sphere can be calculated as the intensity reflected by the sample. Having into account the linear and nonlinear optical contributions, this can be expressed as:

$$\mathcal{R}_N(z,\theta) = \frac{\int_o^\infty |R(\theta)|^2 I(r,z)rdr}{\int_o^\infty |R_0(\theta)|^2 I(r,z)rdr} \quad (2)$$

In equation (2) $z = d_s - f$, $r$ is the spatial coordinate associated to the transversal area of a Gaussian beam, and $I = I_o \left(\frac{w_o}{w}\right)^e \exp\left(-2\frac{r^2}{w^2}\right)$ is the beam intensity on the sample surface. Expanding $R(\theta)$ in a Taylor series: $= \bar{R}_0(\theta) + \Delta \bar{n} \frac{\partial R(\theta)}{\partial \bar{n}}\Big|_{\Delta \bar{n}=0}$, with $\bar{R}_0(\theta)$ being the linear complex reflection coefficient and $\Delta \bar{n} = \bar{n}_2 I$, we can found an expression to $\mathcal{R}_N$:

$$\mathcal{R}_N(f,\theta) = 1 + \frac{Re\{\delta(\theta)\}}{1 + \left(\frac{d_s - f}{z_0}\right)^2} \quad (3)$$

The quantity $\delta(\theta) = \frac{\bar{n}_2 I_0}{R_0} \frac{\partial R(\theta)}{\partial \bar{n}}\Big|_{\Delta \bar{n}=0}$ is defined as the normalized-nonlinear reflection coefficient, and it is the ratio between the nonlinear and the linear contributions on the reflection coefficient at $r = 0$ and $z = 0$. In this way, the final expression for $\mathcal{R}_N$ is:

$$\mathcal{R}_N(f,\theta) = 1 + Re\left\{\left(\frac{2\bar{n}_0^3 \cos\theta - 4\bar{n}_0^2 \sin^2\theta \cos\theta}{\bar{n}_0^4 \cos^2\theta - \bar{n}_0^2 + \sin^2\theta}\right) \cdot \frac{(n_2 + i\kappa_2)I_0}{\sqrt{\bar{n}_0^2 - \sin^2\theta}}\right\} \cdot \frac{1}{1 + \left(\frac{d_s - f}{z_0}\right)^2} \quad (4)$$

Equation (4) is the expression used to fit the experimental data. Figure 5 shows the RF-scans for both perovskites when the incident angle is near to 60°. The obtained values for the nonlinear refractive index are $n_2 = (-0.2 \pm 0.6)\,\text{cm}^2/\text{GW}$ and $n_2 = (-0.12 \pm 0.02)\,\text{cm}^2/\text{GW}$ for 3D and 2D perovskites, respectively. From the measured values of two-photon absorption coefficient and nonlinear refractive index, we can calculate the nonlinear figure of merit[26] $F = n_2/(\beta\lambda)$. This quantity is an important parameter to show that these materials can be used in several photonics applications such as optical limiting devices an all-optical switching devices. The values of nonlinear optical parameters as well as the figure on merit for 3D and 2D hybrid perovskites are listed in Table I.

TABLE I. Nonlinear optical properties for 2D and 3D perovskites.

| Sample | Wavelength (nm) | Thickness (nm) | $\beta$ (cm/MW) | $n_2$ (cm$^2$/GW) | $F$ |
|---|---|---|---|---|---|
| 2D | 1064 | 700 | $-25.6 \pm 7.3$ | $-0.12 \pm 0.02$ | 0.04 |
| 3D | 1064 | 500 | $-6.0 \pm 2.0$ | $-0.2 \pm 0.6$ | 0.31 |

Other important aspect of the results obtained here is the difference between TPA values for 2D and 3D perovskites (around one order of magnitude). It is found that 2D perovskite can give rise to multiple-quantum-well structures, in which the organic part serves as potential wells, and the organic layer functions as potential barriers[10]. The photo-generated bound excitons are therefore confined within the inorganic slab, and when combined with the dielectric screening, a larger excitons binding energy is obtained. The large binding energy can enhance the nonlinear response of the system due to strong oscillation of the generated excitons[27].

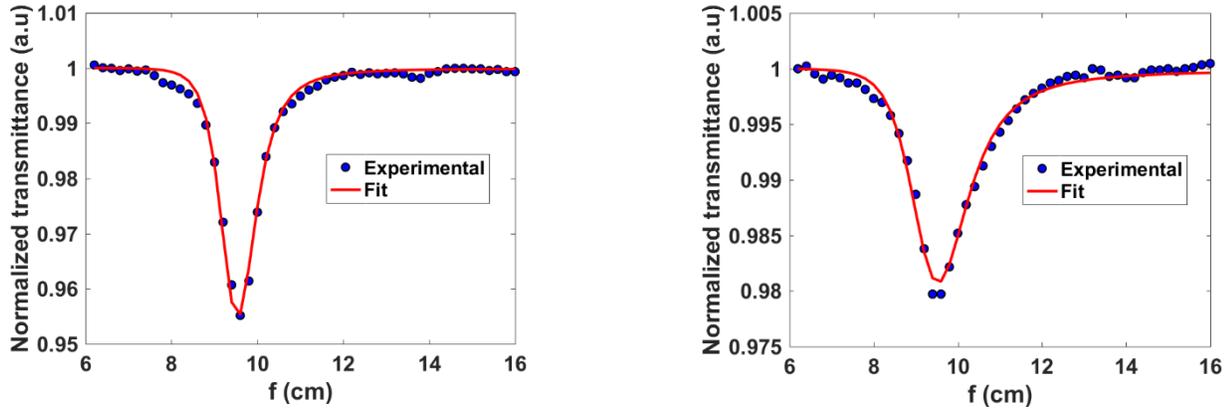

FIG 5. RF-scan results for a) 3D Perovskite and b) 2D Perovskite.

In conclusion, we have measured the TPA coefficient and nonlinear refractive index for hybrid perovskite materials with different dimensionality. We used a TF-scan and RF-scan techniques at 1064 nm for MAPbBr$_3$ and (PA)$_2$(MA)$_2$Pb$_3$Br$_{10}$, referred as 3D and a 2D perovskite thin films, respectively. The values obtained for these parameters are $\beta = (-6.0 \pm 2.0)$ cm/MW and $n_2 = (-0.2 \pm 0.6)$ cm$^2$/GW for MAPbBr$_3$ 3D perovskite and $\beta = (-25.6 \pm 7.3)$ cm/MW and $n_2 = (-0.12 \pm 0.02)$ cm$^2$/GW for (PA)$_2$(MA)$_2$Pb$_3$Br$_{10}$ 2D perovskite. These results predict a nonlinear figure of merit for photonics applications suggesting 3D and 2D hybrid perovskite as potential materials for photonic devices. We compared the dimensionality effect for both materials, and we found that TPA value in 2D hybrid perovskite is one order of magnitude higher that 3D perovskite. This difference is explained if we considered the effect of exciton confinement, which enhances the nonlinear optical response of the system.

D.R, F.J, J.I.U and J.O. Thank the Colombian "Departamento Nacional de Planeación", SGR collaborative project 2013000100184 between Empresas Públicas de Medellin, Andercol S.A., Sumicol S.A.S., and Universidad de Antioquia, for supporting this work. We would like to thank the program "Estrategia de Sostenibilidad 2017-2019 de la Universidad de Antioquia". J. Serna acknowledge the support from Universidad Pontificia Bolivariana. E. Rueda thanks Universidad de Antioquia U de A and H. Garcia thanks Southern Illinois University, Edwardsville for financial support.

[1] S.D. Stranks, G.E. Eperon, G. Grancini, C. Menelaou, M.J.P. Alcocer, T. Leijtens, L.M. Herz, A. Petrozza, and H.J. Snaith, Science (80-. ). **341**, (2014).

[2] T. Leijtens, S.D. Stranks, G.E. Eperon, R. Lindblad, E.M.J. Johansson, J.M. Ball, M.M. Lee, H.J. Snaith, and I.J. Mcpherson, ACS Nano 7147 (2014).

[3] X. Zhao, B. Zhang, R. Zhao, B. Yao, X. Liu, J. Liu, and Z. Xie, J. Phys. Chem. Lett. **7**, 4259 (2016).

[4] J.C. Yu, D. Bin Kim, E.D. Jung, B.R. Lee, and M.H. Song, Nanoscale **8**, 7036 (2016).

[5] D. Ramirez, Y. Suto, N.C. Rosero-Navarro, A. Miura, K. Tadanaga, and F. Jaramillo, Inorg. Chem. **57**, 4181 (2018).

[6] Q. Fu, X. Tang, B. Huang, T. Hu, L. Tan, L. Chen, and Y. Chen, Adv. Sci. (2018).

[7] S. Mirershadi, S. Ahmadi-Kandjani, a. Zawadzka, H. Rouhbakhsh, and B. Sahraoui, Chem. Phys. Lett. **647**, 7 (2016).

[8] W. Liu, J. Xing, J. Zhao, X. Wen, K. Wang, P. Lu, and Q. Xiong, Adv. Opt. Mater. **5**, 1 (2017).

[9] B.S. Kalanoor, L. Gouda, R. Gottesman, S. Tirosh, E. Haltzi, A. Zaban, and Y.R. Tischler, ACS Photonics **3**, 361 (2016).

[10] Y. Chen, Y. Sun, J. Peng, J. Tang, K. Zheng, and Z. Liang, Adv. Mater. **30**, 1 (2018).

[11] D.H. Cao, C.C. Stoumpos, O.K. Farha, J.T. Hupp, and M.G. Kanatzidis, J. Am. Chem. Soc. **137**, 7843 (2015).

[12] H. Tsai, W. Nie, J.C. Blancon, C.C. Stoumpos, R. Asadpour, B. Harutyunyan, A.J. Neukirch, R. Verduzco, J.J. Crochet, S. Tretiak, L. Pedesseau, J. Even, M.A. Alam, G. Gupta, J. Lou, P.M. Ajayan, M.J. Bedzyk, M.G. Kanatzidis, and A.D. Mohite, Nature **536**, 312 (2016).

[13] D. Ramirez, J.I. Uribe, L. Francaviglia, P. Romero-Gomez, A. Fontcuberta i Morral, and F. Jaramillo, J. Mater. Chem. C (2018).

[14] N. Giesbrecht, J. Schlipf, L. Oesinghaus, A. Binek, T. Bein, P. Muller-Buschbaum, and P. Docampo, ACS Energy Lett. acsenergylett.6b00050 (2016).

[15] S. Yakunin, L. Protesescu, F. Krieg, M.I. Bodnarchuk, G. Nedelcu, M. Humer, G. De Luca, M. Fiebig, W. Heiss, and M. V Kovalenko, Nat. Commun. **6**, 8056 (2015).

[16] J.-H. Cha, J.H. Han, W. Yin, C. Park, Y. Park, T.K. Ahn, J.H. Cho, and D.-Y. Jung, J. Phys. Chem. Lett. acs. jpclett.6b02763 (2017).

[17] A. Bernasconi and L. Malavasi, ACS Energy Lett. acsenergylett.7b00139 (2017).

[18] J. Serna, A. Hamad, H. García, and E. Rueda, in *Photonics 2014 12th Int. Conf. Fiber Opt. Photonics* (Kharagpur, India, 2014).

[19] R. Kolkowski and M. Samoc, J. Opt. **16**, 125202 (2014).

[20] J. Serna, J. Henao, E. Rueda, A. Hamad, and H. García, **arXiv:1807.03752v1** **[physics.optics]**.

[21] D. V Petrov, A.S.L. Gomes, and C.B. De Araujo, Appl. Phys. Lett. **65**, 1067 (1994).

[22] M. Martinelli, L. Gomes, and R.J. Horowicz, Appl. Opt. **39**, 6193 (2000).

[23] D. V. Petrov, J. Opt. Soc. Am. B **13**, 1491 (1996).

[24] R.A. Ganeev and A.I. Ryasnyansky, Phys. Status Solidi Appl. Mater. Sci. **202**, 120 (2005).

[25] M. Sheik-Bahae, A. a. Said, T.-H. Wei, D.J. Hagan, and E.W. Van Stryland, IEEE J. Quantum Electron. **26**, 760 (1990).

[26] M. Dinu, F. Quochi, and H. Garcia, Appl. Phys. Lett. **82**, 2954 (2003).

[27] G. Chakravarthy, S.R. Allam, A. Sharan, O.S.N. Ghosh, S. Gayathri, A.K. Viswanath, M.N. Prabhakar, and J.-I. Song, J. Nonlinear Opt. Phys. Mater. **25**, 1650019 (2016).